\documentstyle[aps,twocolumn,epsfig]{revtex}
\begin{document}
\twocolumn[\hsize\textwidth\columnwidth\hsize
\csname@twocolumnfalse%
\endcsname

\draft 

\title{Bose-Einstein condensates in strong electric fields - \\
       effective gauge potentials and rotating states}
\author{J. M. Kailasvuori and T. H. Hansson}
\address{Department of Physics, University of Stockholm,
         Stockholm Center for Physics, Astronomy and Biotechnology, 
         S-11385 Stockholm, Sweden}
\author{G. M. Kavoulakis}
\address{Mathematical Physics, Lund Institute of Technology, P.O. Box 118,
         S-22100 Lund, Sweden}
\date{\today}

\maketitle

\begin{abstract}

Magnetically-trapped atoms in Bose-Einstein condensates  
are spin polarized. Since the magnetic field is inhomogeneous,
the atoms aquire Berry phases of the Aharonov-Bohm type during adiabatic
motion. In the presence of an eletric field there is an additional 
Aharonov-Casher effect. Taking into account the limitations on 
the strength of the electric fields due to the polarizability 
of the atoms, we investigate the extent to which these effects can be
used to induce rotation in a Bose-Einstein condensate.

\end{abstract}
\pacs{PACS numbers: 72.15.R, 03.75.F, 05.30.J, 32.60}

\vskip0.1pc]

\section{Introduction}

The ability to control and manipulate clouds of trapped cold atoms has 
resulted in a series of interesting experiments on Bose-Einstein condensates. 
A very striking example is the creation and observation of quantized 
vortex states, which are expected to occur in rotating atomic condensates 
\cite{dalibard,ketterlev}, much in the same way as in liquid helium II. 

In this paper we study the quantum mechanical effects related to
the adiabatic motion of the spin-polarized atoms in strong external
electromagnetic fields. As will be explained in section II below, 
``adiabatic'' here refers to the dynamics of the spin; spin flip 
transitions are assumed to be suppressed by a large Zeeman gap.

One of our main conclusions is that the magnitude of the electric 
field $E$ required to induce a vortex state is very high and
approximately  given by the simple relation
\begin{eqnarray}
   e E R \sim m_e c^2= 0.5 \, \mathrm{MeV}, 
\label{res}
\end{eqnarray}
where $R$ is the size of the condensate, and $m_e$ is the electron mass. 
Since $E$ is limited both by direct experimental difficulties, and by 
polarization effects of the atoms (that essentially change the trapping 
potential), this condition cannot be satisfied in present experiments. It is, 
however, not excluded that future experiments with much larger condensates
could be in a regime where these effects would become important.
Apart from this possible  application, we hope that the 
theoretical methods used in this paper can be of some general interest.

In the presence of magnetic and electric fields, the atoms aquire quantum
phases during adiabatic motion, and these effects can, as originally shown by
Berry, be represented by effective gauge potentials. The most famous --
and here also the most obvious -- example is the Berry phase due to the 
adiabatic rotation of the spin along with an external magnetic field  
${\bf B} = B\hat{\bf B}$. 
For a closed path $C$ the phase $\gamma_{B}$ aquired is given by
$  \hbar \gamma_{B} =- \hbar m_s  \Omega$
where $\hbar m_s$ is the projection of the total spin along the $z$ axis and
$\Omega$ is the solid angle swept out by ${\bf \hat B}({\bf r})$ along the
curve $C$. As shown by Berry, this phase can also be expressed as a 
line integral of a vector potential, ${\bf a}_B$, or, using Stokes theorem, as 
a surface integral of the corresponding effective magnetic field, ${\bf b}_B$,
\begin{eqnarray}
  \hbar \gamma_{B} =
   \oint_{C} d{\bf r} \cdot {\bf a}_B  =
   \int{d} {\bf S} \cdot {\bf b}_B 
\label{ber}
\end{eqnarray}  
In the following, 
we will refer to ${\bf a}_B$ and ${\bf b}_B$ as the Berry potential
and Berry field, respectively, and to avoid confusion in the notation,
we shall denote real external electromagnetic fields and vector potentials with
capital letters, and effective fields and potentials with small
letters. 

Below we shall incorporate the Berry phase in the description of the 
Bose-Einstein condensate by  coupling a complex scalar condensate wave function to  the vector 
potential ${\bf a}_B$. 
By noting that ${\bf b}_B$ is a monopole field in the parameter space
spanned by ${\bf \hat B}({\bf r})$, it is easy to find the 
corresponding expression in ${\bf r}$ space,
(see, \emph{e.g.}, \cite{girvin} or \cite{leinaas})
\begin{eqnarray}
     ({\bf b}_B)_i \equiv (\nabla \times {\bf a}_B)_i =
    - \frac{1}{2}\hbar  m_s \epsilon_{ijk} (\partial_{j} {\bf \hat B} 
  \times\partial_{k} {\bf \hat B}) \cdot {\bf \hat B}, 
\label{aba}
\end{eqnarray}  
Because of the monopole character of the field the corresponding
vector potential has a string like singularity or, alternatively, is 
given by a nontrival fibre bundle. This complication will be of no 
relevance for this paper since we shall not attempt to explicitly
solve the field equations in the presence of ${\bf a}_B$.

In a geometry which is such that the particles move in a region where 
the effective magnetic field vanishes, there can still be a purely topological 
effect of the Aharonov-Bohm type, but more typically  ${\bf b}_B$  
is nonvanishing throughout space and the particle experiences a
Lorentz force  
(which is much smaller than the ``Stern-Gerlach'' force $\nabla
(\vec{ \mu} \cdot {\bf B})$ responsible for the magnetic trapping of the 
atoms). 

When electric fields are present, new effects occur. Qualitatively, 
this is because an atom  moving with  velocity $\bf v$ 
in an electric field ${\bf E}$, will in its rest frame see 
a magnetic field ${\bf B}^\prime$ given by 
\begin{eqnarray}
  {\bf B}^\prime = {\bf B} - \frac 1 {c^{2}}{\bf v} \times {\bf E} 
  + {\cal  O}(v^2/c^2) \, ,
\label{lorb}
\end{eqnarray} 
and get a Zeeman shift $\propto \vec{ \mu}\cdot {\bf B'}$ 
where $\vec{\mu}$ is the magnetic moment of the spin polarized atom. 
This effect can be incorporated in the Lagrangian 
describing the system by
an effective vector potential 
\begin{eqnarray}
{\bf a}_{AC} \equiv \vec{ \mu} \times
{\bf E}/c^2,
\end{eqnarray}
 which is the origin of the  Aharonov-Casher (AC) effect
\cite{ac}.  Just as in the case of the Berry field strength discussed above, 
the Aharonov-Casher field, ${\bf b}_{AC}$, is nonvanishing 
for a general configuration of the external
electric and magnetic fields. In particular, this can be the case for a
constant $\bf B$ field, as long as $\bf E$ is space dependent, or vice
versa. The latter case is the most interesting for magnetically-trapped atoms.

Effective magnetic fields in Bose-Einstein condensates  can induce
rotations, just as an ordinary ${\bf B}$ field in a type II 
superconductor.  
 However, for realistic geometries the Berry potential ${\bf a}_B$ 
can only give a phase $< 2\pi m_{s}$, and since a single vortex 
corresponds to a $2\pi$ rotation of the 
phase of the condensate wavefunction, this can at most induce a 
couple of vortices in condensates of alkali atoms.
Since the strength of the AC potential is $\propto {\bf E}$, 
it can in principle give large phases, but since 
it is a relativistic effect, this would require very strong electric 
fields. An intersting feature of the AC potential is that the corresponding 
effective magnetic field, ${\bf b}_{AC}$ has a  spatial 
distribution  very different from that of the effective AB magnetic field. 

The physical origin of Berry and Aharonov-Casher phases are well 
understood, and the aim 
of this paper is to investigate whether they are of importance in 
realistic configurations of atomic Bose-Einstein condensates. 
However, to make the presentation more self-contained, 
Sec.\,II contains a short summary of how the effective potentials
${\bf a}_B$ and ${\bf a}_{AC}$ can be derived, and also how
they can be incorporated into the Gross-Pitaevskii equation satisfied by the
condensate wave function. In Sec.\,III we estimate the magnitude of the
induced phases, and the corresponding circulation for various field
configurations, and in particular for a quadrupole trap. In this 
connection we also note that the effects of a topological Aharonov-Casher
phase in a toroidal configuration was studied earlier by Petrosyan and 
You \cite{petyou}.

Since very strong electric fields are required for the effects 
of ${\bf a}_{AC}$ to be non-negligible, it is important to investigate how
electric polarization effects could influence the  potential trapping the 
atoms, and this is done in Sec.\,IV.
Finally, we summarize our conclusions in Sec.\,V.

\section{Effective vector potentials in the Gross-Pitaevskii equation}

We start with the Hamiltonian for a single particle
with mass $m$, charge $q$, spin $s$ and magnetic moment $\vec{\mu}$
in an external electric and magnetic field,
\begin{eqnarray}
  H = \frac {1} {2m} [ {\bf p} - q  {\bf A}({\bf r})
  -{\bf a}_{AC}({\bf r}) 
 ]^{2} - q A_{0}({\bf r})  - \vec{ \mu} \cdot {\bf B}({\bf r}),
\label{pham}
\end{eqnarray}
where $A^\mu=(A^{0},{\bf A})$ is the usual EM gauge field, ${\bf a}_{AC} = 
\vec{\mu} \times {\bf E} /c^2$, and $\vec{ \mu} = \mu {\bf S}$. 
The spin operator $\bf S$ is a matrix acting on
the $(2s+1)$-component wave function.
For a spin 1/2 particle, and to linear order in the fields, this
Hamiltonian follows from the Dirac equation
(see, \emph{e.g.}, \cite{anandan}), and we will simply assume the same form of 
$H$ for particles with general charge and spin, and in particular for neutral 
bosons.

The second-quantized description of a collection of such
fully-polarized neutral bosons with spin $s$   
is given by a coherent-state path integral for the $2s+1$-component 
complex scalar field $\vec\psi$ with a Lagrangian,
\begin{eqnarray}
     {\cal L} &=& \vec\psi^{\dagger} (i \hbar \partial_{t} + \mu 
     B {\bf S})\vec\psi \nonumber \\ &-&
    \frac 1 {2m} 
  \vec\psi^{\dagger}\left(-i \hbar \nabla - {\bf a}_{AC} \right)^{2}\vec\psi 
 -\frac \lambda 4 (\vec\psi^{\dagger}\cdot \vec\psi)^{2}, 
\end{eqnarray}
where we have added a contact-interaction term with strength $\lambda$. 

The next step is to use an adiabatic 
approximation to ``freeze out'' the spin degree of freedom, assuming
that the $\bf B$ field is strong enough to  polarize all the atoms in 
a state given by the magnetic quantum number $m_{s}$. 
Technically, we proceed by decomposing $\vec\psi = 
\psi\vec\chi(\hat {\bf s}) = 
\psi {\bf U}(\hat{\bf s})\vec\chi_{\uparrow}$, where $\psi$ is a 
single component complex scalar field, 
and $\vec\chi(\hat {\bf s})$ a $2s+1$ component 
spinor satisfying $\vec\chi(\hat {\bf s})^{\dagger}\cdot
\vec\chi(\hat {\bf s}) =1$ and 
$\vec\chi(\hat {\bf s})_{m}^{\dagger}{\bf S}\vec\chi(\hat {\bf s})
=\hat{\bf s}$\cite{schulz}. Here $\vec\chi_{\uparrow}$ is a constant spinor 
corresponding to having the spin fully polarized in the, say, 
$z$-direction, and $\bf 
U$ is a unitary operator that rotates $\vec\chi_{\uparrow}$ to the 
space time dependent spinor $\vec\chi(\hat {\bf s}({\bf r},t))$ that 
describes a spin at the position ${\bf r}$ pointing in the direction 
of the unit vector $\hat {\bf s}({\bf r})$.

The adiabatic 
approximation for the spin dynamics now  amounts to fixing this unit vector 
along the local direction of the magnetic field,  {\em i.e.} taking 
$\hat {\bf s} 
({\bf r},t) =\hat{\bf B}({\bf r},t)$, and neglecting fluctuations in the spin 
direction. (Note that we do not assume the orbital motion to be 
adiabatic.) The resulting path integral
\begin{eqnarray}
     {\cal L} = \psi^{\dagger} (i \hbar \partial_{t} + m_s \mu B)\psi 
   &-&  \frac 1 {2m} 
  \psi^{\dagger}(-i \hbar \nabla - {\bf a}_{AC}
  - {\bf a}_B )^{2}\psi   \nonumber   \\
 &-&\frac \lambda 4 (\psi^{\dagger} \psi)^{2}, 
 \label{lag1}
\end{eqnarray}
 is now over the single component field $\psi$, which 
however couples to the effective gauge potential
\begin{eqnarray}
{\bf a}_{B} =i \hbar (U^{\dagger}\nabla U)_{\uparrow\uparrow}.
\label{abu}
\end{eqnarray}
It is easy to verify that the vector potential ${\bf a}_{B}$, is 
nothing but the Berry potential discussed in the introducction.
Also note that (for $m_{s}>0$) the  term $m_s \mu B$ acts as a
trapping potential for atoms.  In the case of a time dependent ${\bf 
B}$, there is also a time component, ${\bf a}^{0}_{B}$, in the Berry 
potential, which would be relevant {\em e.g.} in a TOP trap. 
We again stress that the gauge potentials emerge because of the  
adiabatic assumption, which in this context amounts to ignoring
spin-flip processes (which are suppressed by the Zeeman gap, assumed
to be large enough.) 

The saddle point of the action of Eq.\,(\ref{lag1}) is
determined by the Gross-Pietaevskii equation 
\begin{eqnarray}
   i \hbar \partial_{t} \phi = \left[ \frac {\hbar^2} {2m} 
  (-i \hbar \nabla - {\bf a}_{\rm eff})^{2} - m_s\mu B 
 -\frac \lambda 2 (\phi^{\dagger} \phi) \right]\phi, 
 \label{gp}
\end{eqnarray}
where $\phi$ is the condensate wave function, and ${\bf a}_{\rm eff} = 
{\bf a}_{B} + {\bf a}_{AC}$. 

We note that Eq.\,(\ref{gp}) resembles the time-dependent 
Ginzburg-Landau equation describing a superconductor, in the sense that there 
is a complex condensate wave function coupled  to a space-dependent gauge 
potential. From this analogy we would expect tha  an 
effective magnetic field would either be expelled, or penetrate in the form of 
quantized vortices. We shall elaborate on this in the next section.

\section{Effective potentials, currents and circulation}

To understand the effects of the effective vector potentials ${\bf a}_B$ and 
${\bf a}_{AC}$, we follow the corresponding analysis for a superconductor. 
There is, however, an important difference in that (at least in the 
approximation considered here) the gauge field is not dynamical but 
rather a background field determined by the orbital motion. In 
particular there are no kinetic terms, and thus no 
reason to minimze the action with respect to the gauge potentials. 
>From this follows that there can be no screening of the effective
Berry or Aharonov-Casher fields, and consequently no Meissner effect. 
We thus expect our system to act as an extreme type II 
superconductor. 
(Only if kinetic terms together with 
${\bf b}_B^{2}$ or ${\bf b}_{AC}^{2}$ terms were generated by 
{\em e.g.} a renormalization group procedure,  could there be a Meissner effect.)
In this sense our system is also very similar to a bucket of helium II 
where rotation with an angular frequency ${\bf \omega}$ can be 
described by a minimal coupling to an effective gauge field
${\bf b}_{\rm eff} = 2 m \vec{ \omega}$. 
In our case, however, the effective field can have a 
more general space time dependence, just as in a superconductor. 

With these comments, we now proceed to minimize the ground state 
energy for a fixed effective magnetic field, and as in the cases of a 
superconductor or a rotating Bose-Einstein condensate, the essence of the 
argument is related to the single-valuedness of the phase of $\phi$. 
(For simplicity we consider only the zero temperature case. It is 
not hard to generalize to finite temperatures well below the Zeeman 
energy by introducing a suitable free energy.)

For a homogeneous system, we parametrize $\phi = \sqrt{\rho_{0}} 
e^{iS/\hbar}$, 
with $\rho_0$ being the constant mean atomic density, and 
 ${\bf j} = \rho_{0} {\bf v} = \rho_{0} (\nabla S - {\bf a}_{\rm eff})/m$
 the corresponding current density.
The single-valuedness of the wave function then gives the usual 
condition $\oint d {\bf r} \cdot (m {\bf v} + {\bf a}_{\rm eff}) = n h$
on the circulation of the velocity field. For a cylindrically-symmetric
geometry, and a circular 
path of radius $r$ around the $z$ axis, we get
\begin{eqnarray}
   \oint  \,r \, d\theta  [ mv_{\theta}(z,r) 
  + ({\bf a}_B + {\bf a}_{AC})_{\theta} ]   =  n h.
\label{circ}
\end{eqnarray}

As a first  illustration, let us assume ${\bf a}_{\rm eff}(r,\theta,z) 
=a_{\rm eff}(r){\hat{\theta}}$ and take as an ansatz solution 
a  vortex  along the $z$ axis with vorticity $n$.  Equation 
(\ref{circ}) then implies
\begin{eqnarray}
    m r v(r) = {n \hbar - \frac{1}{2 \pi} {\Phi_{\rm eff}}(r)},
\label{vel}
\end{eqnarray}
where $\Phi_{\rm eff}(r) = {\Phi_{AC}} + {\Phi_{B}}$ is the flux due to
the Aharonov-Casher and the Berry magnetic fields through the surface spanned
by a circle of radius $r$.
The kinetic energy per unit length $\epsilon (n)$ associated with the vortex 
state is then given by
\begin{eqnarray}
    \epsilon(n) &=& \int\, d^2r \frac 1 2 m \rho(r) v^2  
\nonumber \\
     &\propto& \int_\xi^R  \frac {dr} {r} 
  [nh -\Phi_{\rm eff}(r)]^2 ,
\label{vort}
\end{eqnarray}
where $\xi$ is the healing length and $\rho(r)$ is approximated by
$\rho_0$ for $\xi < r < R$, and taken as  zero elsewhere.

Just as a rotating bucket of Helium, or a type II superconductor, 
we expect the presence of the magnetic flux 
to induce rotation in the form of vortex states. To get a rough estimate of
when this happens, we assume a constant density and ${\Phi_{\rm eff}}(r) =
kr^l$. One can then show that for
vortices to be energetically favourable, \emph{i.e.}, $\epsilon (n) < \epsilon (0)$,
one must have 
\begin{equation}
   \oint_{r=R} d{\bf r} \cdot {\bf a}_{\rm eff} =
 {\Phi_{\rm eff}}(R) \sim n h,
\label{neweq}
\end{equation}
where $R$ is the radius of the condensate.
Neglecting the Berry phase, this is nothing more 
than the condition that the Aharonov-Casher phase be of order 1,
in agreement with the estimate of Ref.\,\cite{yiyou}, where a toroidal 
geometry was used and a purely topological Aharonov-Casher phase was considered.
The general case with non-constant density must be treated by solving 
the Gross-Pitaevskii equation, which can be done numerically, 
but we will not do it in the present study.

Next we investigate the field configurations needed to achieve 
${\Phi_{\rm eff}}\sim h$. As a first illustrative 
(but completely unrealistic) 
example, we consider a constant magnetic field ${\bf B} = B_{0} {\bf \hat z}$,
and an electric field with a constant gradient, ${\bf E} = ({E_{0}} y/R) 
{\bf \hat x}$, implying ${\bf a}_B = 0$ and ${\bf a}_{AC} = (\mu E_{0}/R) 
{\bf \hat y}$. Therefore, the effective magnetic field is constant, 
${\bf b}_{AC} = (\mu E_{0}/R) {\bf \hat z}$. Thus we have ${\Phi_{\rm eff}}  
= \pi R^2 b_{AC} = \pi R \mu E_{0}/c^2$, which yields the condition
\begin{equation}
   e E_0 R \sim m_e c^2.
\label{e:fundam}
\end{equation}

We now turn to a more realistic example, and assume that the 
atoms are trapped in a quadrupole field ${\bf B}= B^\prime(x,y,-2z)$ 
by being in a state of polarization
\begin{equation}
    \vec {\mu} = -\mu \frac{x {\bf \hat{x}} + y {\bf \hat{y}} - 
  2 z {\bf \hat{z}}}{\sqrt{x^2+y^2+4z^2}},
\end{equation}
\emph{i.e.}, opposite to the polarizing field ${\bf B}$. This state is the so-called 
low-field seeking state (the absence of trapping at the center of the trap
is a problem, since the atoms which are located there escape from the
trap, and there are numerous tricks for fixing that, like in TOP traps or 
Ioffe-Pritchard traps; we do not worry about these extra complications 
in the present study.) 

\begin{figure}
\begin{center}
  \begin{small}
    \epsfig{file=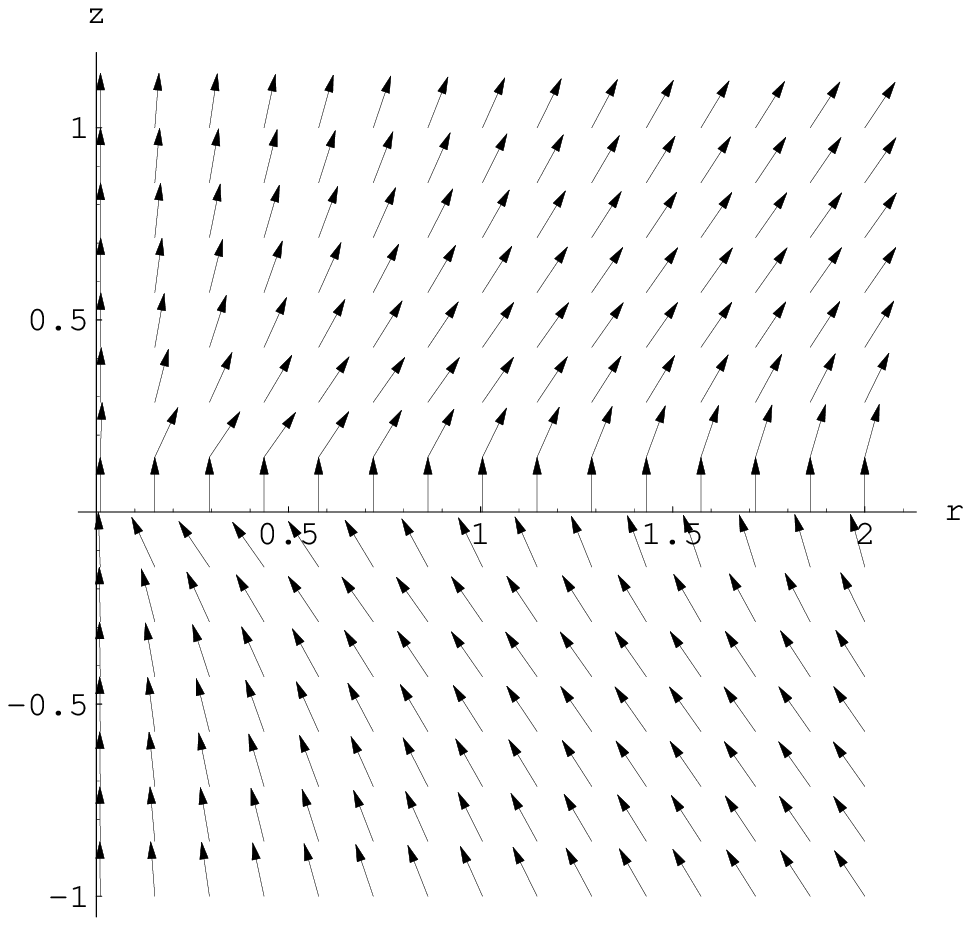,width=0.32\textwidth}
    \epsfig{file=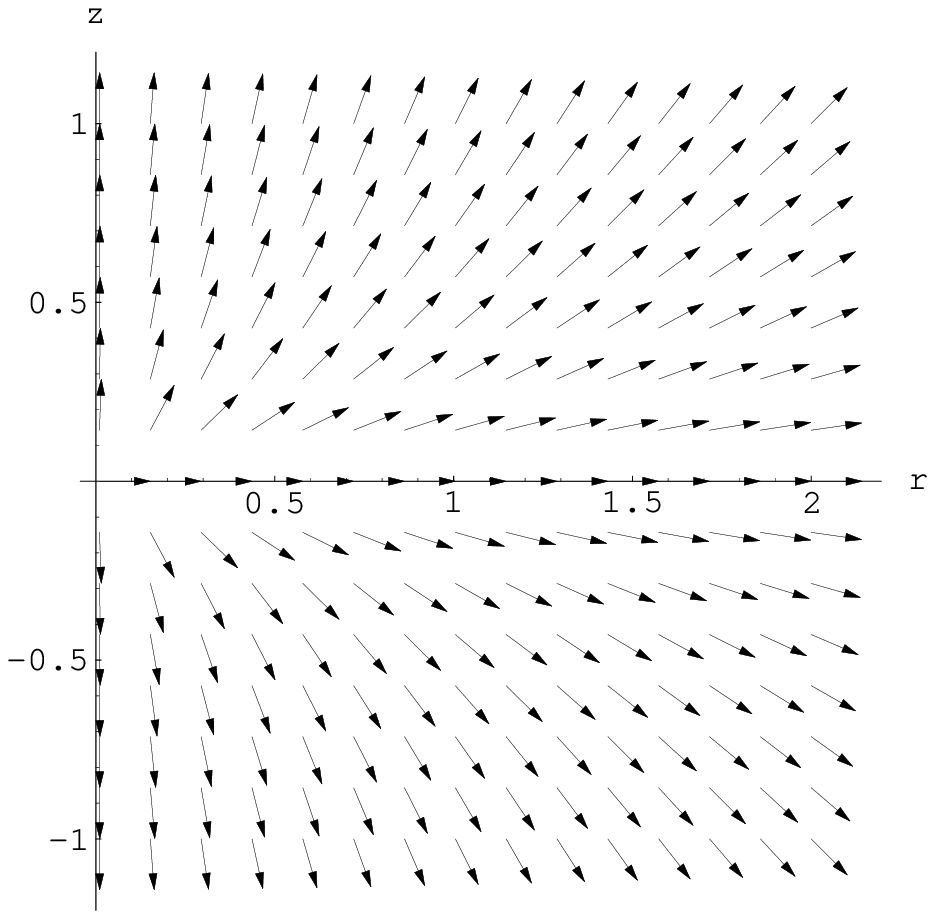,width=0.32\textwidth}
  \end{small}
\end{center}
\caption{The Aharonov-Casher field (higher) and the Berry field (lower)
in polar coordinates. The vectors are normalized to be of constant length.}
\label{f:abcplots}
\end{figure}

With ${\bf E}=E_0 {\bf \hat z}$ we have (at $r=R$, \emph{i.e.}, in the 
periphery of the cloud),
\begin{eqnarray}
     {\bf a}_{AC} &=& \frac {1} {c^2} {\vec \mu} 
    \times {\bf E} = \frac{\mu E_0 R}{c^2 \sqrt{R^2+4z^2}}
   {\bf \hat{\theta}}, 
\end{eqnarray}
and 
\begin{eqnarray}
  {\Phi_{AC}} &=& 2\pi \mu E_0 \frac{R^2}{\sqrt{R^2+4z^2}}.
\label{qac}
\end{eqnarray}
The polarization also adds to ${\bf a}_{\rm eff}$ the contribution from 
the Berry potential Eq.\,(\ref{abu}):
\begin{eqnarray}
    {\bf a}_B &=& \hbar m_s(\cos \varphi -1) {\bf \hat{\theta}},   
\end{eqnarray}
\begin{eqnarray}
   {\Phi_{B}} &=& h m_s (\cos \varphi -1) = - 
  h m_s \left(\frac{2z}{\sqrt{R^2+4z^2}}+1 \right), 
\label{qab}
\end{eqnarray} 
where $\cos \varphi = {\bf \hat{B}} \cdot {\bf \hat{z}}= -2z / \sqrt{R^2+z^2}$.
This expression for the Berry potential is derived within a spin one-half 
representation by using the unitary operator $U=e^{-i \phi {\bf \hat m}
\cdot {\vec \sigma}/2}$, where ${\bf \hat m}= {\bf \hat B} \times {\bf \hat z}/ 
|{\bf \hat B} \times {\bf \hat z}|$ describes the rotation axis,
and $\varphi = \arccos ({\bf \hat B} \cdot {\bf \hat z})$ 
is the rotation angle. This potential has a Dirac string singularity
on the positive $z$ axis. In Fig.\,\ref{f:abcplots}
we illustrate the corresponding fields. 

Since the particles are bosons, $m_s$ is an integer and the term $-h m_s$ in
${\Phi_{B}}$ corresponds to an integer number of flux quanta and induces no
vorticity. The remaining part vanishes when integrated over the volume
occupied by the entire cloud. Therefore, the Berry potential induces no
vorticity.

For $z=0 $, ${\Phi_{AC}}(R) = 2 \pi R \mu E_0/c^2$, and we retain the result
of Eq.\,(\ref{e:fundam}) obtained above.
In fact, this result holds for general flux configurations
of the form ${\Phi_{\rm eff}}(r)\propto r^l$.
The knowledge of the maximal electric field strength that can be applied to 
the condensate yields the minimum size of the cloud for our method of
inducing rotation to be applicable.

\section{Limits on the electric fields due to the atom polarizability.}

Application of an electric field results in a Stark shift $\Delta$, which 
in the quadratic regime can be thought of as the interaction
between the electric field and an induced electric dipole moment, 
which is  
\begin{equation}
  \Delta = - {\bf d} \cdot {\bf E} = - \frac{1}{2}\alpha E^2.
\label{e:polar}
\end{equation}
Here $\alpha$ is the polarizability of the atoms, and ${\bf d}= 
\alpha {\bf E}/2$ is 
the electric dipole moment. For alkali atoms $\alpha/\hbar \sim 100\,
\rm{kHz/(kV/cm)^2}$. The induced dipole moment changes
the physics of the trapped atoms in 
several ways. The most direct effect is that the Stark shift 
changes the ground-state energy so that spatially-varying $\bf E$ 
fields change the trap geometry. In addition, the dipole-dipole 
interaction given by
\begin{eqnarray}
  V_{\rm{dd}}({\bf R}) &=& \frac{ {\bf d}_1 \cdot {\bf d}_2 -
    3({\bf d}_1 \cdot {\bf \hat{R}})({\bf d}_2 \cdot {\bf \hat{R}})}
  {4 \pi \epsilon R^3}, 
\label{e:dipp}
\end{eqnarray}
where 
${\bf R}$ is the distance between two atoms, and $\epsilon$ is the 
permittivity, 
induces spin-flip transitions of the
atoms into non-trapped states so they can escape. 
The dipole-dipole interactions do not conserve the spin but 
the magnetic dipole-dipole interaction is sufficiently 
small (compared to the Coulomb interaction) for the spin 
relaxation rate to be low enough for the condensate to be observable.

We have not attempted to calculate these quite complicated effects, 
but we made some estimates based on the following simple considerations. 
First, notice that the interaction given by Eq.\,(\ref{e:dipp}) differs 
from the magnetic dipole-dipole interaction only by a constant factor.
In vacuum the interactions are of the same order when $d / \sqrt{ \epsilon_0} \sim
\mu \sqrt{\mu_0}$. For these values the 
spin relaxation rate should therefore be equally
small. In the regime of the quadratic Stark effect,
$d = \alpha E/2$, with $\alpha/\hbar = 100\,{\rm kHz}/({\rm kV/cm})^2$,
which implies $E\sim 10^7\,{\rm V/m}$.

A more thorough study of the electric dipole-dipole 
interaction can be found in \cite{yiyou},
where an effective scattering length is defined based on the
effective potential
\begin{equation}
   V_{\rm{eff}}({\bf R}) = u_0\delta ({\bf R})+ V_{\rm{dd}}({\bf R})
\label{e:kaisav}
\end{equation}
where $u_0=4\pi \hbar^2 a_{\rm sc}/m$, with $a_{\rm sc}$ being the
scattering length for atom-atom elastic collisions described by the 
contact-interaction term in Eq.(\ref{e:kaisav}).
For ${}^{87}\mathrm{Rb}$, the effective scattering length corresponding to
$V_{\rm{eff}}({\bf R})$ is relatively unaffected up to $E \sim 10^{7}\,
{\rm V/m}$. For higher values it changes drastically, attaining a negative
value for
\begin{equation}
  E \approx 6 \times 10^7\,{\rm V/m}.
\label{e:yyb}
\end{equation}
This value actually provides an upper bound for $E$. We therefore estimate
$E \sim 10^7\,{\rm V/m}$ to be the upper limit for the
field strength. Inserting this into Eq.\,(\ref{e:fundam}) yields 
that the lowest possible value of $R$ is $\sim 5$ cm.                     

\section{Summary}

We have demonstrated that inhomogeneous electric and magnetic fields
applied in trapped atomic Bose-Einstein condensates can  in principle be used
in order to induce vortex states. More specifically, we have inferred  
the corresponding Gross-Pitaevskii equation satisfied
by the condensate wave function in the presence of an electric
and a magnetic field, and we have investigated the influence of effective 
vector potentials such as the Berry and  Aharonov-Casher type
on the condensate wavefunction. The Aharonov-Casher field can be used to
induce vorticity in the cloud and we have found that the magnitude of the
electric field $E$ and the condensate size $R$ have to satisfy the equation 
$e E R \sim m_e c^2= 0.5$ MeV. As a relativistic effect, it requires large
electric fields for typical sizes of the condensate. The
magnitude of the electric field is in turn limited by the polarizability of
the atoms. Estimating these limitations we find a
lower limit for the condensate size to be $R \agt 5$ cm.
 
Finally we want to stress that the mere possibility to study gauge
interaction in Bose-Einstein condensates is of considerable interest.
For instance, we know that condensed systems in the presence of
magnetic fields can exhibit different Meissner phases. As already
indicated, this could be of relevance if the effective gauge fields somehow
would aquire dynamics.
Two dimensional fermi systems in strong magnetic fields exhibit
various quantum Hall phases, and with this in mind it would be interesting
to apply the methods developed in this paper both to {\em e.g.}
lower dimensional atomic Bose-Einstein condensates and
degenerate  fermion systems.

\vskip2pc
We thank S. Stenholm for useful discussions and comments on the manuscript.
J.K. also wishes to acknowledge A. Kastberg,  E. Lindroth and C.J. Pethick 
for useful discussions. G.M.K. was supported financially by the 
Swedish Research Council (VR),
and by the Swedish Foundation for Strategic Research (SSF).

\end{document}